%% file: main.tex
\documentclass[conference]{IEEEtran}
\IEEEoverridecommandlockouts

\usepackage{eso-pic}
\usepackage{lmodern}
\usepackage{gensymb}

\usepackage{color}
\usepackage{xspace}
\usepackage{amsmath}
\usepackage{url}
\usepackage{booktabs}
\usepackage{multirow}
\usepackage{graphicx}
\usepackage{caption}
\usepackage{soul}
\usepackage[british]{babel}
\usepackage[normalem]{ulem}
\usepackage{makecell}
\usepackage{pifont}
\usepackage{subfigure}
\usepackage[absolute]{textpos}

\newcommand{\cmark}{\ding{51}}
\newcommand{\xmark}{\ding{55}}

\newcommand{\BULLET}{\vspace{+.00in} \noindent $\bullet$ \hspace{+.00in}}

\newcommand{\etc}{\emph{etc.}\xspace}
\newcommand{\ie}{\emph{i.e.,}\xspace}
\newcommand{\eg}{\emph{e.g.,}\xspace}
\newcommand{\etal}{\emph{et al.}\xspace}

\newcommand{\fka}{\emph{f.k.a.}\xspace}

\newcommand{\todo}[1]{{\color{red}[\textsf{TODO: #1}]}}
\newcommand{\csq}[1]{{\color{green}#1}}
\newcommand{\rz}[1]{{\color{black}#1}}
\newcommand{\note}[1]{{\color{cyan}}}
\newcommand{\rev}[1]{{\color{black}#1}}

\newcommand{\accessdate}{25-July-2022}

\title{\huge Will Metaverse be NextG Internet? Vision, Hype, and Reality}

\author{\IEEEauthorblockN{Ruizhi Cheng, Nan Wu, Songqing Chen, and Bo Han}
\IEEEauthorblockA{
George Mason University\\
Email: \{rcheng4,nwu5,sqchen,bohan\}@gmu.edu}
}
\begin{document}

\begin{textblock}{8}(5,0.3)
\noindent\Large \textcolor{red}{Published in the IEEE Network\\
\url{https://ieeexplore.ieee.org/document/9877927}
}
\end{textblock}

\maketitle

\begin{abstract}

Metaverse, with the combination of the prefix ``meta'' (meaning transcending) and the word ``universe'', has been deemed as the next-generation (NextG) Internet.
It aims to create a shared virtual space that connects all virtual worlds via the Internet, where users, represented as digital avatars, can communicate and collaborate as if they are in the physical world. 
Nevertheless, there is still no unified definition of the Metaverse. 
\rz{This article first reviews what has been heavily advocated by the industry and the positions of various high-tech companies.
It then presents our vision of what the key requirements of Metaverse should be.} \note{(Comment 1 from Reviewer 4)}
After that, it briefly introduces existing social virtual reality (VR) platforms that can be viewed as early prototypes of Metaverse and conducts a reality check by diving into the network operation and performance of two representative platforms, Workrooms from Meta and AltspaceVR from Microsoft.
Finally, it concludes by discussing several opportunities and future directions for further innovation.

\end{abstract}

\vspace{0.1in}

\iffalse
\fi

\section{Introduction}
\label{sec:intro}
\input{01.intro}

\section{Industry Trends}
\label{sec:industry}
\input{03.industry}

\section{Defining Metaverse}
\label{sec:definition}
\input{02.definition}

\section{Social VR Platforms}
\label{sec:socialVR}
\input{04.socialVR}

\section{Case Studies}
\label{sec:case-studies}
\input{05.case-studies}

\section{Discussion}
\label{sec:discussion}

\input{06.discussion}

\section{Conclusion}
\label{sec:conclusion}
\input{07.conclusion}

\bibliographystyle{unsrt}
\bibliography{bib/metaverse,bib/s&p}

\end{document}

%% file: 01.intro.tex
%\bo{%consider drawing a figure to illustrate the big picture of the metaverse? 
%consider replacing metaverse with Metaverse?}

%\cite{park2022metaverse} %https://ieeexplore.ieee.org/document/9667507

Although the term %basic concept of
Metaverse has been around for almost 30 years since it was coined by %American writer 
Neal Stephenson in his 1992 science fiction novel {\em Snow Crash}, we are still in the early stage of actually building the Metaverse, which envisions an immersive successor to the Internet. 
The development of Metaverse has gone through several stages.
Retrospectively, 
%In the initial stage (the late 1970s), %\csq{but the term was coined in 1992? } 
the text-based interactive games, such as MUD (multi-user dungeon) that emerged in the late 1970s, could be viewed as the earliest prototypes of Metaverse, even before the term was literally introduced.
They define a multiplayer %real-time
virtual world with role playing, interactive fiction, and online chat.
The second phase happened during the postmillennial decade with the development of commercial virtual worlds such as Second Life\footnote{\url{https://secondlife.com/} (accessed on \accessdate)}. %, which is also considered as the first metaverse prototype~\cite{dionisio2013virtual}.
%
%The third stage started around 2007 and 
It then embraced fully 3D virtual worlds such as OpenSimulator\footnote{\url{http://opensimulator.org/} (accessed on \accessdate)}, which is largely compatible with Second Life.

In the current stage, with the flourishing of 5G and mobile immersive computing~\cite{han2019mobile}, there has been a surge of research \& development on the Metaverse in both industry and academia.
We have now entered an open
development phase of the Metaverse, which is widely considered as a %combination of Internet technologies  (such as 5G), 
collection of 3D virtual worlds connected via the Internet~\cite{dionisio2013virtual} and enabled by various immersive technologies such as augmented reality (AR), virtual reality (VR), and mixed reality (MR), which are often collectively referred to as extended reality (XR). 
%
%\bo{Internet technologies provide users high quality of experience (QoE) and low latency network support, while XR gives users an immersive experience.} 
%
%\crz{Internet technologies and XR work together to provide users with high Quality of Experience (QoE) network support and immersive experiences.}
%
While there is still no unified definition of the Metaverse, % (Section~\ref{sec:definition}), 
\rz{it is broadly deemed as a hypothetical next-generation (NextG) Internet\footnote{\url{https://bit.ly/3cn5SCr} (accessed on \accessdate)}.} \note{(Comment 6 from Reviewer 4: Add a footnote.)}

\begin{figure}[t]
    \small
    \centering
    \includegraphics[width=0.8\columnwidth]{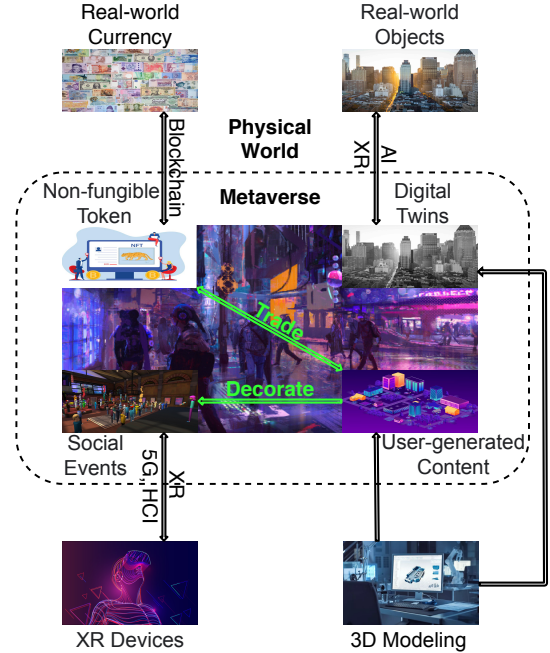}
    \vspace{-0.1in}
    \caption{\small Elements of the Metaverse and their interaction with the physical world.
    %\bo{replace the three appearances of AR/VR with XR}
    %\bo{potential copyright issues?}
    }
    \label{fig:metaverse}
    \vspace{-0.2in}
\end{figure}

Figure~\ref{fig:metaverse} illustrates the basic elements in the Metaverse and how they interact with the physical world.
In general, users with XR devices access the Metaverse and participate in its various social events, whose smooth execution is enabled by techniques such as 5G and HCI (human-computer interaction).
%
%The Metaverse server will use communication technologies such as 5G, as well as human-computer interaction (HCI) technologies to support the smooth running of the Metaverse.
%
Users are free to create their own content via 3D modeling to decorate social events in the Metaverse.
The content can be traded using non-fungible tokens (NFTs) through a decentralized blockchain. % technology. % that corresponds in value to the physical world's currency.
Physical objects can be presented in the Metaverse as digital twins that are generated via 3D modeling and consumed with XR devices assisted by artificial intelligence (AI).
%
%and , while using technologies such as AI to ensure that real-world objects and the digital twin interact and influence each other.

%\crz{Should we unify the "real-world" and the "physical world"?}
%\bo{describing Figure~\ref{fig:metaverse}}

In this article, we present our vision of the Metaverse by discussing its key technical requirements (Section~\ref{sec:definition}). 
We then review recent advances in the industry and introduce the advocates of various key players %high-tech companies
(Section~\ref{sec:industry}). 
After that, we provide an overview of existing social VR platforms, the early prototype of Metaverse that combines online social networks and VR technologies, and compare their unique features (Section~\ref{sec:socialVR}). 
We then conduct a first-of-its-kind reality check to understand the networking protocol usage and system performance of two representative platforms, Meta's Horizon Workrooms\footnote{\url{https://www.oculus.com/workrooms/} (accessed on \accessdate)} (referred to as Workrooms) and Microsoft's AltspaceVR\footnote{\url{https://altvr.com/} (accessed on \accessdate)} (Section~\ref{sec:case-studies}).
Finally, we discuss the technical challenges, opportunities, and directions for future research activities (Section~\ref{sec:discussion}) and conclude this article.

%% file: 03.industry.tex
% \bo{What is happening in the industry?  Key aspects and challenges of metaverse?} \crz{How to cite news and data source? }

% \bo{add a figure to summarize the main effort of different companies?}
\begin{figure}[t]
    \small
    \centering
    \includegraphics[width=0.99\columnwidth]{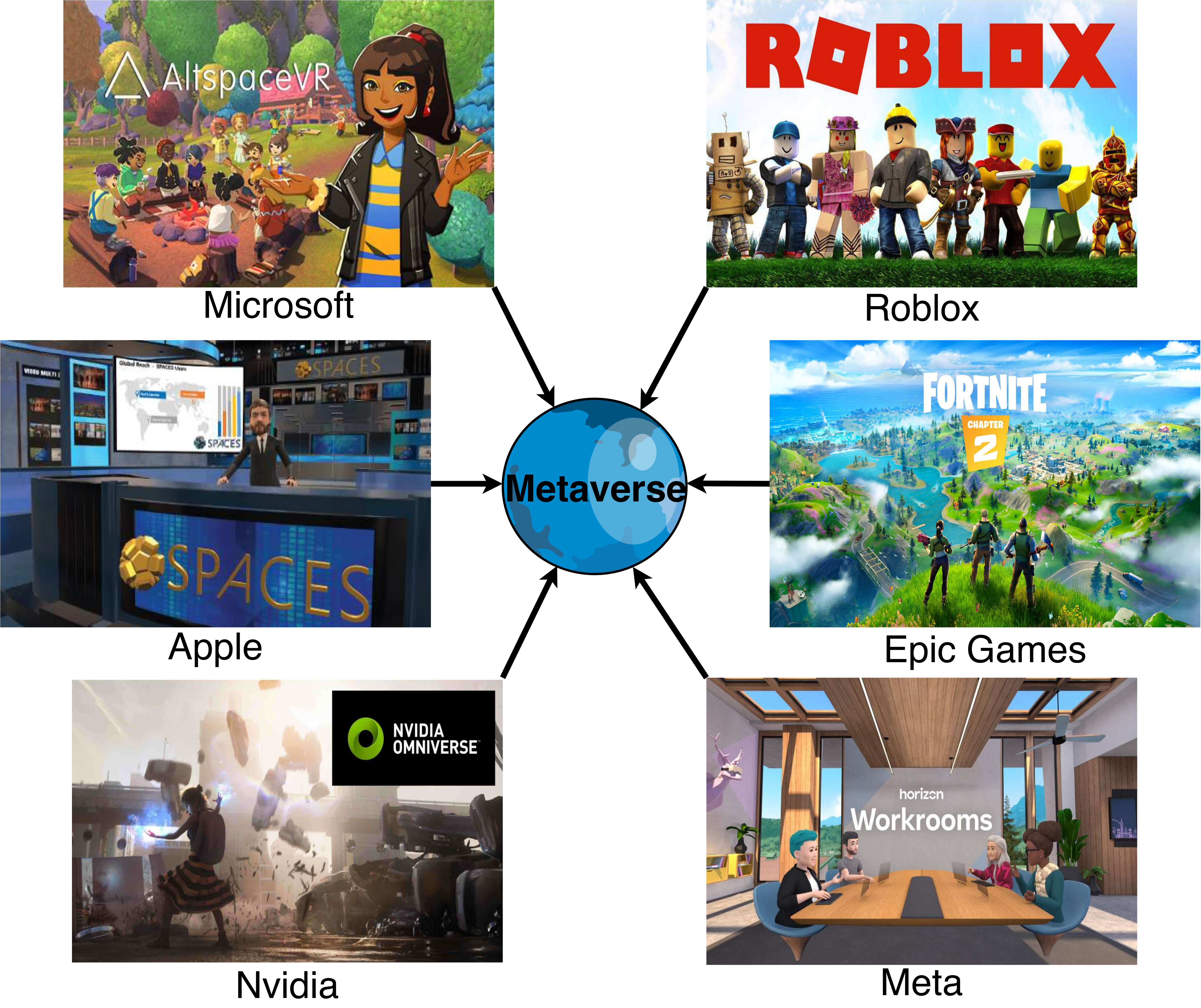}
    \vspace{-0.1in}
    \caption{\small Current development of the Metaverse in industry.
    }
    \label{fig:industry}
    \vspace{-0.2in}
\end{figure}

In this section, we briefly introduce the current development of Metaverse in the industry, which is summarized in Figure~\ref{fig:industry}.

%We then present the key challenges of building the Metaverse from the network perspective.

%\vspace{0.05in}
%\noindent {\bf Current Industry Development.}
%In recent years, with the booming development of 5G and immersive computing technologiies such as AR/VR/MR, Metaverse is back on the horizon of investors. 
Many high-tech companies have joined the Metaverse arena.
Meta is conceivably the most notable among all %companies
that have invested in this space. 
In September 2019, Meta (named Facebook then) announced \textit{Facebook Horizon}, a social VR platform. 
In July 2021, Facebook announced the transition into a Metaverse company within five years. 
To echo this vision, in October 2021, Facebook changed its name to Meta. 
%
%Meta has invested \$10+ billion to build the Metaverse in 2021 alone and will continue to increase the investment in the coming years. 
%
Meta considers VR as the foundation to build the Metaverse.
%, which may be related to their VR products. 
%
Its VR headset, Oculus Quest 2, has sold over 10 million units, making it the state-of-the-art and best-selling VR device. % in the world. 

Nvidia %, on the other hand, 
%a GPU manufacturer, 
announced a plan to create the first virtual collaboration and simulation platform called Omniverse\footnote{\url{https://www.nvidia.com/en-us/omniverse/} (accessed on \accessdate)} in August 2021. 
This platform can be used to connect 3D worlds into a shared virtual universe and create digital twins, simulating real-world buildings and factories. 
%
%\crz{
Omniverse has three key components. %can be divided into three parts.
The first one is Omniverse Nucleus, a database engine that allows multiple users to connect and create a scene together.
The second one is the rendering and animation engine %used 
to simulate the virtual world.
%
%\bo{
The third one is Nvidia CloudXR for streaming XR content to client devices. %the edge server
%}. %, such as edge server. 
%\crz{``CloudXR is NVIDIA's solution for streaming virtual reality (VR), augmented reality (AR), and mixed reality (MR) content from any OpenVR XR application on a remote server—cloud, data center, or edge''}
%
Meanwhile, Omniverse integrates AI %technology 
to train digital twins in the Metaverse. % within Omniverse, which in turn affects real-world objects.
%}
%\bo{expand this paragraph a bit?}

Epic Games, a video game company famous for its Unreal game engine, announced %to invest 
a \$1 billion investment to build the Metaverse.
In its most popular game, \textit{Fortnite}\footnote{\url{https://www.epicgames.com/fortnite/en-US/home} (accessed on \accessdate)},
%
%\note{footnote?}\csq{footnote of the url or reference are both fine}\note{I have added the footnote for Fortnite and Roblox in the ``Discussion'' section. Do we need to add the footnote in the both place? }
which is regarded as a prototype of the Metaverse, users can create their avatars, buy digital items, and enjoy movies and concerts. % as their avatars. 
Roblox\footnote{\url{https://www.roblox.com/} (accessed on \accessdate)} is another company in this arena. 
As the largest UGC game platform, players in Roblox can create their own games and virtual worlds. 
They can buy, sell, and create virtual items that can be used to decorate their avatars. % on the platform. 
%
%Roblox also supports VR devices to improve users' immersive experience. %The CEO of Roblox wants to expand the Roblox universe into a place where players can learn, work create and socialize. 
%
%Roblox is currently one of the "worlds" that is closest to the Metaverse. 
%However, Roblox mainly targets game players and provides limited XR support. 
%
%\bo{what is the difference between Metaverse and Metaverse?}

%\csq{comment out the last two sentences. shall we measure roblox?}\crz{According to my understanding, roblox is a game platform similar to Steam, but unlike steam, it gives players a lot of freedom. So it's rich in UGC, digitial twins, and other Metaverse elements. But it is mainly aimed at PC, mobile, and game consoles, and currently has limited support for AR/VR.}

%{\bf Economic Consideration}
Although most companies embrace the Metaverse's concepts and vision, 
%consider VR as the cornerstone or main body of building the Metaverse, there are 
%some companies that hold a different view, such as
cautions and doubts also emerge. 
While both Apple and Microsoft have virtual space applications\footnote{Apple acquired a VR company, Spaces, in 2020, and Microsoft acquired a social VR platform called AltspaceVR back in 2017.}, they consider that seamlessly connecting the Metaverse and the physical world is a key to its success, if not more important than the Metaverse itself. 
%
%\crz{
They believe that the %initial 
purpose of creating the virtual space is just to enable people to improve productivity and reduce production costs in the physical world.

%% file: 02.definition.tex
% https://www.theverge.com/22701104/metaverse-explained-fortnite-roblox-facebook-horizon

% https://about.fb.com/news/2021/09/building-the-metaverse-responsibly/

% https://www.trtworld.com/life/introducing-the-next-generation-of-the-internet-the-metaverse-51117

% https://www.washingtonpost.com/technology/2021/08/30/what-is-the-metaverse/

\vspace{0.05in}
\noindent {\bf Existing Definitions and Enabling Technologies.}
Metaverse has been viewed as a new type of %social Internet application, social platform, 
online social network, and arguably NextG Internet. 
%\bo{add more existing definitions of the metaverse?}
%
%For example, it could 
It would be the convergence of digital second life (for ``escape'') and virtual reality (for exploration), mimicking user interaction in the real world.
A narrow definition of Metaverse is thus a universal virtual world focusing on social interaction, which connects multiple 3D virtual environments via the Internet (\ie a network of virtual worlds~\cite{dionisio2013virtual}).
We envision that the Metaverse should evolve to {\em seamlessly integrate the physical world and the virtual space}, for example, via digital twin and digital economies (\eg cryptocurrencies).

%To illustrate it more simply, it is a shared online space that combines 3D graphics on 2D screens (PC, smartphones) or in virtual reality~\cite{sparkes2021metaverse}.
% \crz{I cannot directly access this citation from \url{https://doi.org/10.1016/S0262-4079(21)01450-0}. But I found this article in \url{https://bit.ly/3rLyrxp}}

Objects in the physical world can interact with the Metaverse. %, presented as digital twins. 
%
%An object in the physical world 
They can generate their digital twins through %technologies such as 
3D modeling and keep their digital twins presenting the same state as what is happening in the real world. % through sensors and other devices.
Conversely, after the digital twin is manipulated/processed in the Metaverse, its physical-world state will be changed accordingly.
For example, BMW %, a world-renowned automobile company, 
has used Omniverse\footnote{https://www.nvidia.com/en-us/omniverse/ (accessed on \accessdate)} from Nvidia to construct a fully functional %, real-time  
digital twin of its automobile factory,  
%
%It can simulate large-scale production and finite scheduling with constraints, 
reducing manufacturing costs and increasing productivity.

\begin{figure}[t]
    \small
    \centering
    \includegraphics[width=0.85\columnwidth]{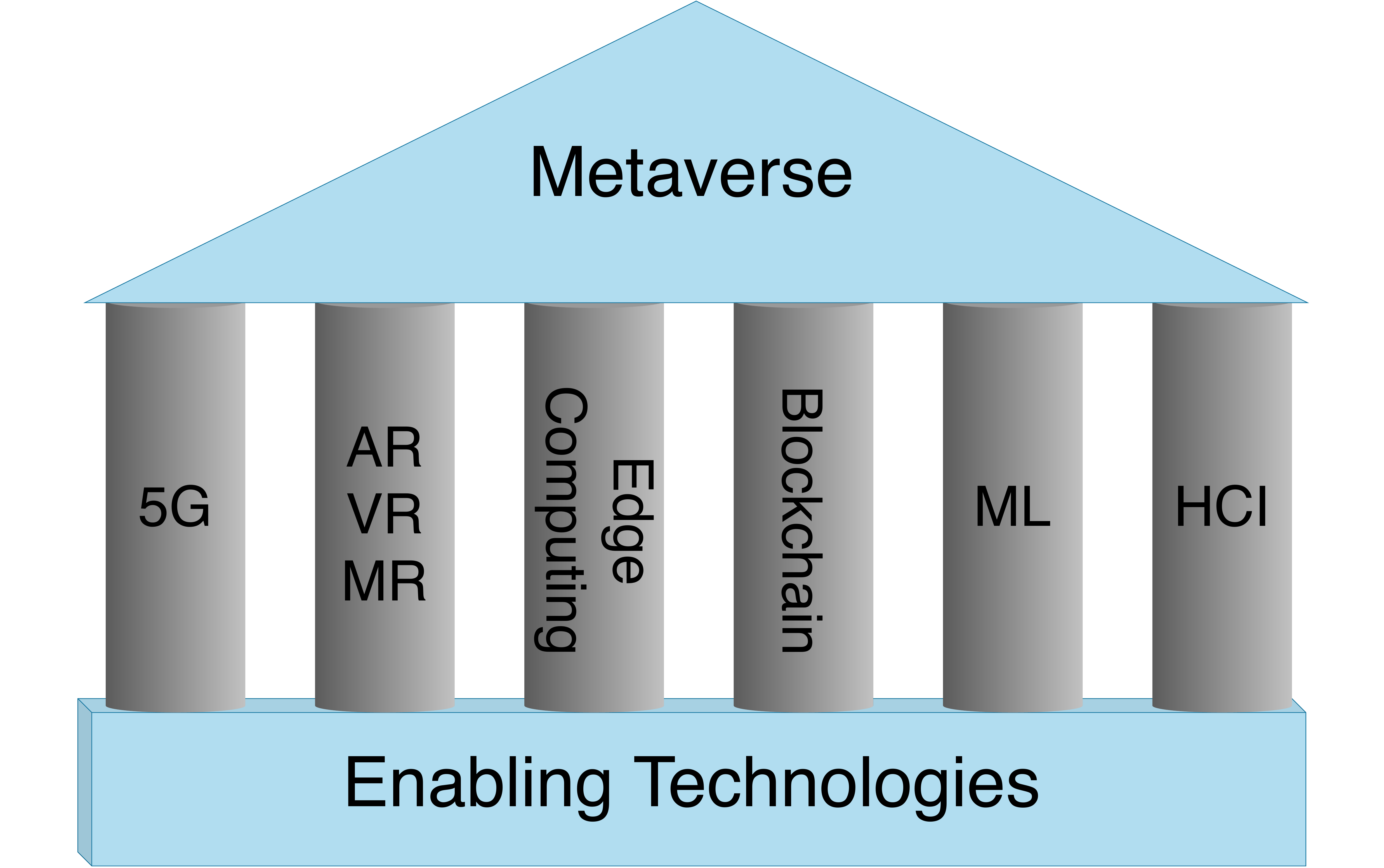}
    %\vspace{-0.1in}
    \caption{\small Enabling Technologies of the Metaverse.
    %\bo{potential copyright issues?}
    }
    \label{fig:pillars}
    \vspace{-0.1in}
\end{figure}
% \bo{draw a figure like the 5G Pillars in https://www.semiconductorforu.com/the-advent-of-5g/?}

\rz{While there is no consensus on the definition, as shown in Figure~\ref{fig:pillars}, it is commonly agreed that the  Metaverse is built on and integrates technologies such as 5G, XR, edge computing, blockchain, %cryptocurrencies,
machine learning (ML), %/artificial intelligence, 
and HCI~\cite{park2022metaverse}.} \note{(Comment 7 from Reviewer 3: Add the reference.)}
%\bo{make sure we discuss all these enabling technologies in the following.}
%
%Metaverse aims to provide users immersive experience based on AR, VR, and MR. %augmented reality (AR), virtual reality (VR), and mixed reality (MR). 

\iffalse
\crz{
Predictably, the Metaverse poses a huge challenge to the network.
%
On the one hand, maintaining the Metaverse, such as accessing massive metadata generated by the sensors, providing social activities and shared virtual space for users, and transmitting high-resolution video streams, requires huge network throughtput. 
%
On the other hand, network latency is crucial to the quality of experience (QoE). 
%
Low latency is particularly critical to motion sickness.
%
Given the users in the Metaverse access from different parts of the world, ensuring low latency when users are across geographically distributed regions is a practical challenge.
%
To provide such a high-throughput, low-latency network, the Metaverse needs to rely heavily on novel network technologies, such as 5G.
%
5G is the state-of-the-art technology standard for cellular networks. 
%
It can reach a maximum throughput of, in theory, 10–20 Gbps.
%
However, considering the scalability demand, the bandwidth requirements of the Metaverse may exceed what 5G can offer~\cite{lee2021metaverse}.
%
}
\fi

%\crz{
\BULLET \textit{5G} %is the fifth generation and state-of-the-art standard for cellular networks. 
%
%It 
provides a faster, lower latency, and more scalable network than 4G.
According to the frequency bands, 5G can be divided into low-band (below 1 GHz), mid-band (between 1 and 6 GHz), and high-band (millimeter-wave, mmWave, from 24 to 39 GHz). 
Low-band 5G %uses frequencies below 1 GHz, similar to those of 4G, but it can provide faster download speeds (about 30-75 Mbps) than 4G.
%
%It 
is used for extensive coverage and is ideal for deployment in rural areas. %\footnote{\url{https://www.t-mobile.com/business/resources/articles/benefits-of-the-5g-spectrum-for-businesses} (accessed on \accessdate)}.
Mid-band 5G %uses frequencies spanning from 1Ghz and 6Ghz, providing download speeds of 100-900 Mbps.
%
%It 
has been commonly deployed in metropolitan areas. % since 2020. %\footnote{\url{https://en.wikipedia.org/wiki/5G} (accessed on \accessdate)}.
High-band 5G %uses millimeter-wave (mmWave) frequency bands (24Ghz - 39Ghz).
%
%It 
can reach a maximum throughput of, in theory, 10-20 Gbps. 
%\bo{this is not consistent with 20 Gbps you mentioned later on!}
%
However, it works in only a small radius, and thus is more useful in urban areas and crowded locations (\eg shopping malls).
%
%High-band 5G had been launched in several U.S. cities, such as Washington D.C., Las Vegas, and New York City. % and so on.
%
%While some carriers believe that mmWare 5G is the real 5G, %some 
%reports demonstrate that mmWare 5G is too expensive to be deployed %and used on 
%at a large scale\footnote{\url{https://venturebeat.com/2019/12/10/the-definitive-guide-to-5g-low-mid-and-high-band-speeds/} (accessed on \accessdate)}.

\BULLET \textit{AR/VR/MR} %are technologies that 
augment or supplant our view of the world, and are a key to the success of Metaverse~\cite{park2022metaverse}.
VR immerses people in the virtual world, % and separated from the physical world.
%
%VR devices, such as VR headsets, provide the usual technical approach to user interaction, such as head tracking and controllers.
%
%VR, especially 
and social VR is widely considered an important component of the Metaverse.
%
%Social VR allows people to communicate, share, and collaborate in a shared virtual world in real time.
%
%Unlike VR, AR provides alternated experiences for users in the physical world. %, which is primarily designed to enhance our physical world.
%
%Compared to VR, AR can be applied to more kinds of devices, such as AR headsets, tabletops, projectors, and so on
%
%Nevertheless, 
%AR headsets have more advantages than other AR devices such as smartphones and tablets, as they allow users to keep their hands away from devices and make them focus more on the augmentation content. % on AR devices.
%
%AR devices interact with people in richer ways, including gestures, vision, sound, and so on.
%
AR enables digital twins in the Metaverse to be overlaid on physical objects in a %n ordinary and
perceptible way, effectively connecting the Metaverse with the physical world.
MR %is a combination of AR and VR that 
allows users to interact with virtual objects, %in a physical environment.
%
%It is often viewed as a more powerful version of AR, 
%It creates 
by creating more connections and collaborative relationships among the physical world, virtual space, and users.

%https://dl.acm.org/doi/10.1145/3290605.3300767

\BULLET \textit{Edge Computing} is a computing paradigm that moves computation and data storage closer to users. %, for example, to cellular base stations. % or WiFi access points. %is a form of computing that takes place near the request data source, reducing the latency experienced by users.
%
%Many technology companies believe 
%It is widely believed that 
The advantageous performance of edge computing in reducing latency for XR %in virtual worlds 
has made it an important backbone for building the Metaverse.
%
%For example, 
Several %worldwide 
telecom carriers have undertaken a project called HoloVerse to test the best 5G edge network infrastructure %around the world 
for efficient deployment of services in the Metaverse\footnote{\url{https://yhoo.it/3AD6dsu} (accessed on \accessdate)}.
Meanwhile, Niantic, the producer of $Pok\Acute{e}mon Go$, has joined forces with telecom carriers to explore how 5G edge computing can enhance the quality of experience (QoE) for AR games\footnote{\url{https://bit.ly/3L1Uj0r} (accessed on \accessdate)}.

\BULLET \textit{Blockchain} %is a distributed database stored in digital form and shared via nodes of the computer network.
%
%The innovation of blockchain is that it 
ensures the security of data records and generates trust without requiring trusted third parties.
%
%It is most well-known for its key role in cryptocurrencies, such as Bitcoin, for the maintenance of secure and decentralized transaction records.
%
It is closely related to user-generated content (UGC) such as digital assets that can greatly enrich the Metaverse~\cite{ryskeldiev2018distributed}. 
For example, NFT, which is used for trading in the Metaverse, is a data unit on the blockchain.
%
%Blockchain will also directly act on other data interactions in the metaverse, such as data storage, data sharing, and so on.
%
%\crz{
%
%However, since UGCs in the metaverse are freely generated by users, they need to be define ownership.
Defining the ownership of UGC in the Metaverse is a practical challenge, as digital assets can be copied and reproduced. 
NFT provides an effective way to prove that UGC is unique and non-fungible (\ie non-interchangeable). % in the Metaverse.
%
%More specifically, users can store their UGC as the NFT on the blockchain, and trade them through smart contracts.
%
It enables owners of digital content to sell/trade their property via smart contracts in the decentralized crypto space (\eg using cryptocurrencies), % based on blockchain. for example, by using cryptocurrencies.
%
%As a concept introduced in 2014, NFT has been growing extremely fast in recent years. 
%
%In 2021, its market value has reached more than \$40 billion\footnote{\url{https://www.ft.com/content/e95f5ac2-0476-41f4-abd4-8a99faa7737d} (accessed on \accessdate)}.

\BULLET \textit{Machine Learning}, especially deep learning (DL), is an important branch of artificial intelligence (AI) that enables machines to learn %and improve performance 
from massive amounts of data.
%
%Regression, classification, clustering, and dimensionality reduction are the main tasks of ML.
%
%Deep learning (DL) is the hottest research area of machine learning in recent years. %, and it is guided by biological neural networks. 
%
%It often requires more data than traditional machine learning algorithms to achieve satisfactory performance.
%
Undoubtedly, the Metaverse will %inevitably
generate huge amounts of complex data, providing rich opportunities for DL. %the application of ML and DL.
%
%ML and DL can analyze and learn useful data and patterns from the massive amounts of data to help make the metaverse and the physical world better.
%
For example, we can use digital twins in the Metaverse for intelligent healthcare.
Laaki~\etal~\cite{laaki2019prototyping} designed a prototype for remote surgery using digital twins %created for the
of patients.
Surgical operations %performed by the doctor
on the digital twin will be repeated on the patient using a robotic arm assisted by DL.
%
%The prototype also uses DL techniques for intelligent patient diagnosis and health prediction.

\BULLET \textit{HCI} focuses on the interaction between users and computers. 
%
%In the metaverse, HCI is mainly used to study how users interact with the digital twins.
%
%\bo{
Given that the final stage of the Metaverse will interconnect the physical world and digital twins, %making the physical and virtual worlds constantly interact with each other.} 
%\crz{Given that the final stage of the Metaverse will be the interconnection of the physical world and its digital twin, so that the physical and virtual worlds constantly interact with each other.}
%
it is, therefore, necessary to enable users to interact with digital twins in real-time and in multiple ways.
The most important problem %that needs 
to be addressed is user input. %, and the most commonly used input devices are %physical devices such as 
%mice, keyboards, and controllers. 
%
The key limitation of existing input devices (\eg mice and keyboards) %such as mice, keyboards, and controllers 
is that they cannot free the users' hands and accurately reflect their body movements.
Recently, researchers have begun to study freehand manipulation that allows more intuitive and concrete interaction %with objects 
in the Metaverse.
These techniques often rely on computer vision and brain-computer interfaces. % (CV) technology.
%
%In addition to freehand manipulation, smart weaving is another emerging user input technique.
%
%It integrates novel materials and conductive threads into common fabrics to support user interactions. % with user interfaces (UI).
%
%Technology companies have begun to invest in this technology, for example, the Jacquard project of Google\footnote{\url{https://atap.google.com/jacquard/}  (accessed on \accessdate)}.
%}

%~\cite{NFT}.
%}
%Users can interact with objects in the physical world based on their digital twins, and \bo{switch between} the real world and the virtual world via some economic systems (\eg  a blockchain based one). 
%Metaverse remains an evolving concept, with different participants enriching its content based on their own understanding.

%\bo{briefly talk about NFT?}

\begin{table*}[t]
\small
\center
\begin{tabular}{c|cccccccc}
%\toprule 
Platforms & Company & Quest 2 & Facial & Game & Personal & Share & Shopping & NFT\\
 & & & Expression & \rev{Events} & Space &  Screen &  & \\
%\midrule
\hline  
AltspaceVR ('15)& Microsoft&\cmark&\xmark&\cmark&\cmark&\cmark&\xmark& \xmark\\ \hline

Bigscreen ('16)&BigScreen&\cmark&\xmark&\xmark&\xmark&\cmark&\xmark& \xmark\\\hline

Rec Room ('16)&\makecell[c]{Rec Room}&\cmark&\cmark&\cmark&\cmark&\xmark&\cmark&\cmark\\ \hline

Anyland ('16)& Anyland&\xmark&\xmark&\cmark&\xmark&\xmark&\xmark& \xmark\\ \hline

VRChat ('17)&VRChat&\cmark&\cmark&\cmark&\cmark&\xmark&\xmark&\xmark\\ \hline

Cluster ('17)&Cluster&\cmark&\cmark&\cmark&\xmark&\xmark&\xmark& \xmark\\\hline

Hubs ('18)&Mozilla&\cmark&\xmark&\xmark&\xmark&\cmark&\xmark& \xmark\\\hline

Workrooms ('21)& \makecell[c]{Meta}&\cmark&\cmark&\xmark&\xmark&\cmark&\xmark&\xmark\\

%\bottomrule 
\end{tabular}
%\vspace{-0.05in}
\caption{\small Comparison of several important features offered by eight social VR platforms. \rev{Personal space is a protective zone in the virtual environment that users can define.}}
\vspace{-0.1in}

\label{tab:social-VR comparision}
\end{table*}

\vspace{0.05in}
\rz{\noindent {\bf Our Vision on Technical Requirements.}} \note{(Comment 1 from Reviewer 4)}
Next, we present our vision of the Metaverse by illustrating three key requirements on scalability, accessibility, and security, privacy, and legal issues.

\iffalse
\csq{
+: text->image+video->3D

+: passive -> active :input; more UGC 

==> trend is a flat and more distributed network

?: accessibility: one is the interfacing; the other is to access anytime anywhere (or availability): same world to all ppl, or different worlds for different classes of ppl?

?: blockchain: NFT

?: security and legal:

?: scalability
}
\fi

%\csq{
{\em Requirement \#1: Scalability.}
With the Internet transitioning to the Metaverse, we expect the first practical challenge faced by any Metaverse platform is the scalability issue. 
As our preliminary investigation in Section~\ref{sec:case-studies} shows, currently Workrooms, an early prototype of the Metaverse, can hardly scale up to %accommodate 
tens of participants. 
When more participants access Workrooms, the corresponding uploading and downloading demand increases proportionally. 
The platform, either serving just as a relay or performing further content processing in the middle, will quickly become a bottleneck.

As can be expected, the bandwidth requirement of Metaverse could be huge. 
On the one hand, compared to traditional 2D videos, the bandwidth for transmitting up to 16K 360-degree panorama~\cite{zhang2021deepvista} or 3D volumetric content~\cite{han2020vivo} to XR headsets could be high.
On the other hand, the Metaverse is full of social elements, which further increases the bandwidth requirement.
Currently, the U.S. Federal Communications Commission (FCC) defines the standard broadband service as 25 Mbps in downlink and 3 Mbps in uplink~\cite{MacMillan2021VCAs}.
Therefore, it is of the utmost importance to guarantee the scalability of Metaverse by leveraging advanced networking techniques.

%}

%\csq{
{\em Requirement \#2: Accessibility.} 
Today's Internet access does not need specific devices.
%
%Any computer (\eg mobile, embedded, or desktop) with a wired or wireless network interface, can freely surf the Internet. 
%
For the Metaverse, however, %as of today, 
users are required to wear headsets for better interaction in the virtual world. 
It greatly limits the accessibility of the Metaverse, %not because the price of such devices, but 
mainly due to the inconvenience of such access. 
%
%To this end, 
We envision that in the future, new ``interfacing'' devices should be developed for accessing the Metaverse without wearing any additional device, and glasses or contact lenses would replace the cumbersome headsets~\cite{park2022metaverse}. 
%
%Ideally, a user desires not to wear any additional device. 
%
%Along this line, Google Starline\footnote{\url{https://blog.google/technology/research/project-starline/} (accessed on \accessdate)} is another promising approach. 
%
Moreover, interaction techniques, other than just display would need to be in place so that users can not only see in the virtual world but also feel, smell, taste, \etc, like what we do in the physical world~\cite{dionisio2013virtual}. 

Besides the interfacing devices of the Metaverse, %is just one potential roadblock. 
another potential obstacle is network accessibility. 
The average 25 Mbps downlink bandwidth in the U.S.~\cite{MacMillan2021VCAs} is far from the demand %that can support 
of even a rudimentary Metaverse -- the bandwidth %and latency
requirement would go up with more and more user-generated content and assets in the Metaverse. 
%
%5G deployment can help relieve such a demand surge. 
%
%More development on the NextG networks would be key. %Data compression, caching, and other techniques could also help. 
%}
Yet another issue related to accessibility is the interoperability across different implementations of the Metaverse, especially when users move from one platform to another.
The user experience should be seamless without any interruption.

%\csq{
{\em Requirement \#3: Security, Privacy, and Legal Issues.} 
Similar to online social networks, in the Metaverse, there will be security and privacy issues, %. This includes the security and privacy issues, 
such as attacks on user authentication and impersonation~\cite{casey2019immersive}. %, that exist in the current Internet, but also new types of security issues, 
\rz{Meanwhile, users’ personal information (\eg biometric data) may be collected for authentication, compromising their privacy~\cite{mathis2021fast}.
Moreover, there will be new types of challenges, for example, securing the NFTs and digital twins, which involve interaction with the physical world. %when they are functioning.
}\note{(Comment 4 from Reviewer 4)}
%buy and trade assets. 
%
%In the Metaverse, there are new types of "assets" that users buy and trade. This involves also the physical world. 
%Some of the discussions advocate blockchain-based solutions to deal with these issues~\cite{xxx}. https://dl.acm.org/doi/10.1145/3474355 
%
%However, blockchains, such as bitcoin\footnote{\url{https://bitcoin.org/} (accessed on \accessdate)} and ethereum\footnote{\url{https://ethereum.org/} (accessed on \accessdate)}, themselves suffer from various attacks. 
%
Moreover, online harassment can be exacerbated in the immersive environment of Metaverse by features including free avatar movements and enhanced feelings of presence and embodiment~\cite{blackwell2019harassment}.
%https://dl.acm.org/doi/10.1145/3359202
%
Furthermore, given that the Metaverse assets (content) are user-generated, there will be copyright issues. 
The protection of content ownership, the detection of copyright infringement, and the licensing of such content have not been well laid out. 
Considering that there will be multiple Metaverse platforms, %to be developed, 
transferring users' assets from one to another is a practical issue to be addressed. 
Such portability and interoperability demand not only standardizations from the industry but also legal enforcement.

%}
%\csq{if space allows, maybe we can discuss a bit more on interoperability/standardization.}

%% file: 04.socialVR.tex
% \bo{Describe different commercial social VR platforms and the key differences.
% Can summarize the key findings from the CS795 report(s).}
% \crz{Should I cite these platforms? Or just give a link in the footnote. TODO}
% \csq{a footnote link is fine. (either way works)}

%\crz{-World +Anyland}

Since social VR is a major component of %building block of 
Metaverse, %and there are already a few social VR platforms, 
we provide an overview of several commercial social VR platforms, %in this section, 
highlighting their key features and differences.
%
%Admittedly, these platforms are in the different development stages towards a real Metaverse. 
%
Social VR, regarded as the future of social media, allows users to interact with each other as avatars in the virtual world, communicating and collaborating %with others 
as if they are in the physical world. 
With the global outbreak of the COVID-19 pandemic, many people around the world have to stay at home and lack social interactions, leading to surging demand for novel applications of social media. 
Thus, predictably, the demand for social VR will continue to grow, as it not only satisfies people's social needs but also gives them a sense of spatial presence.

\iffalse
\begin{table}[t]
\small
\center
\begin{tabular}{c|cccc}
%\toprule 
Platforms & Browser&  Smartphone
& \makecell{PC\\App} & \makecell{ Open \\Source} \\
%\midrule
\hline  
VRChat &\xmark& \xmark&\cmark&\xmark\\ %\hline

Rec Room &\xmark& \cmark& \cmark&\xmark\\ %\hline

AltspaceVR & \xmark& \xmark& \cmark&\xmark\\ %\hline

Mozilla Hubs & \cmark&--%\rule[1.0pt]{0.2cm}{0.2em}
&\xmark& \cmark\\%\hline

Anyland &\xmark& \xmark& \cmark&\xmark\\ %\hline

Cluster &\xmark& \cmark& \cmark&\xmark\\ %\hline

Bigscreen &\xmark& \xmark& \cmark&\xmark\\ %\hline

Workrooms &\cmark&\xmark& \xmark&\xmark\\
%\bottomrule 
\end{tabular}
\vspace{-0.05in}
\caption{\small Comparison of popular social VR platforms.}
\vspace{-0.15in}
% \bo{replace XSolidBold with something else? \cmark and \xmark ~change the order and add more (launch year)? do not delete the old table, comment it out instead.}}

\label{tab:social-VR comparision}
\end{table}
\fi

\vspace{0.05in}
\noindent {\bf Key Features.} %Overview summary.}
After an extensive survey, we focus on the eight %of the 
most popular social VR platforms, %namely
VRChat\footnote{\url{https://hello.vrchat.com/} (accessed on \accessdate)}, Rec Room\footnote{\url{https://recroom.com/} (accessed on \accessdate)}, AltspaceVR, Mozilla Hubs\footnote{\url{https://hubs.mozilla.com/} (accessed on \accessdate)}, Anyland\footnote{\url{http://anyland.com/} (accessed on \accessdate)}, Cluster\footnote{\url{https://cluster.mu/} (accessed on \accessdate)}, Bigscreen\footnote{\url{https://www.bigscreenvr.com/} (accessed on \accessdate)}, and Workrooms. %\bo{what is the order here?}
\rz{As a first step, % of understanding, 
we mainly focus on examining their features and support concerning the accessibility, social events, harassment-protection, entertainment, and interactivity with the real world of these platforms as follows:
%them from the following perspectives: 
%count the following metrics separately: 
\romannumeral1) Whether they are accessible from the popular Oculus Quest 2? \romannumeral2) Whether their avatars have facial expressions? \romannumeral3) Whether they have the personal space feature, which is a zone to protect users from harassment? \romannumeral4) Whether they have the gaming, sharing PC screen, and shopping features, and \romannumeral5) Whether they support the trading of assets with NFTs? } %\note{(Comment 2 from Reviewer 3 and Comment 2 from Reviewer 4)}

\rz{Table~\ref{tab:social-VR comparision} presents a summary of these platforms.
%
% We find that most commercial social VR platforms support most VR headsets, but Meta's social VR platforms (Horizon Worlds and Workrooms) currently only support Meta's headsets, namely the Oculus. And except for Meta's platforms all other platforms provide access options for Windows users. Among these platforms, Altspace VR is the only one that provides a desktop application specifically for macOS users. 
% \csq{you meant except? Hubs does not seem to support windows either? considered Linux users?}\crz{Hubs and Workrooms can be used in PC OS (Windows, macOS, Linux, etc.) through a web browser, while "windows" and "macOS" for VRChat, Rec Room, and Altspace VR refer to their desktop clients. But I find it crowded to put all these things in a table}, \csq{just use check like browser column} all other platforms provide access options for Windows users. Among these platforms, Altspace VR is the only one that provides a desktop application specifically for macOS users. 
%\crz{
We find that all platforms except Anyland support Oculus Quest 2.
Avatar's face expression, game events, personal space, and PC screen sharing are supported by about only half of the platforms, showing varied design choices and development stages across them.
Finally, only Rec Room offers shopping and NFTs, demonstrating that virtual trading is not yet widely available on social VR platforms.
} \note{(Comment 2 from Reviewer 3 and Comment 2 from Reviewer 4)}
\vspace{0.05in}
\noindent {\bf User Experience.}
We experiment with the above %commercial social VR 
platforms and highlight %summarize 
their advantages in terms of QoE. %user experience.
%\crz{How to make the layout more appropriate?}

\BULLET \textit{AltspaceVR:} 
The ambient lighting of the virtual scene matches the shadows, making the lighting of the scene realistic. There are many environments and live events initiated from all over the world, with a rich social element.

\BULLET \textit{Bigscreen:} 
Users can play PC games in the virtual world and watch videos together (\eg Netflix and YouTube) played on PCs in a private or public room.

\BULLET \textit{Rec Room:} It offers an abundance of game activities and enables cross-play between different users with VR headsets, PCs, and smartphones.
%The interaction between users with different devices is smooth.

\BULLET \textit{Anyland:} 
It is a ``sandbox universe'', where users can build anything (even the avatar) they need using tools that exist in the physical world.

\BULLET \textit{VRChat:} Users can build their own games in the virtual world. It allows an impressive amount of customization (\eg users can upload any 3D model as the avatar).

\BULLET \textit{Cluster:} 
%\bo{It supports many devices, including VR headsets, Windows, macOS, Android, and iOS. is this already known from the above description?}\crz{yes, may be need to delete it} \bo{replace it with something else in one sentence}\crz{
% It bears a large capacity for public events, with up to 500 people participating in a four-hour event at the same time, %}
% %
% and allows people to organize events for free.
It has a variety of highly interactive
social events, such as live concerts. In addition, it
can make souvenir books with the photos taken
by users at each event.

\BULLET \textit{Mozilla Hubs:} %The only open-source social VR platform, 
Users can customize their %social VR 
applications with its source code and deploy their own servers. %For PC and smartphone users, 
They can use Hubs through browsers without downloading any application, which is lightweight and convenient.

 %Users can use many tools to create a game for their friends to explore together. 

 %, and pictures of the events can be saved and made into a book.

 %, such as Twitch, Netflix, and YouTube. 
%
%They can customize the size of the virtual screen. 
%
%Multiple users can gather together to enjoy a common experience in a private or public room.

% The game format is %very 
% \bo{creative} and the commands are %very 
% intuitive.

\BULLET \textit{Workrooms:} 
It %allows the user to use a 
supports physical keyboards, which are much more convenient than the virtual ones manipulated by a controller.
%}
Moreover, users can write using the controller as a pen %in a new way 
by flipping it around. % and writing like a pen. 

\noindent {\bf Other Types of Metaverse Platforms.}
\rz{In addition to social VR,  recent advancements of the Metaverse embody massively multiplayer online games, such as Fortnite, Minecraft\footnote{\url{https://www.minecraft.net/en-us} (accessed on \accessdate)}, and Roblox,
as well as the emerging NFT or blockchain-based online games, such as Decentraland\footnote{\url{https://decentraland.org/} (accessed on \accessdate)}, Upland\footnote{\url{https://www.upland.me/} (accessed on \accessdate)}, and Axie Infinity\footnote{\url{https://axieinfinity.com/} (accessed on \accessdate)}.
However, since they are designed primarily for PC users with 2D content, these games currently cannot provide their users with an immersive experience, which is one of the most important goals of the Metaverse.
Hence, in this paper, we focus on the investigation of social VR platforms.}
\note{(Comment 6 from Reviewer 3)}

%% file: 05.case-studies.tex
\begin{figure}[t]
    \small
    \centering
    \includegraphics[width=0.99\columnwidth]{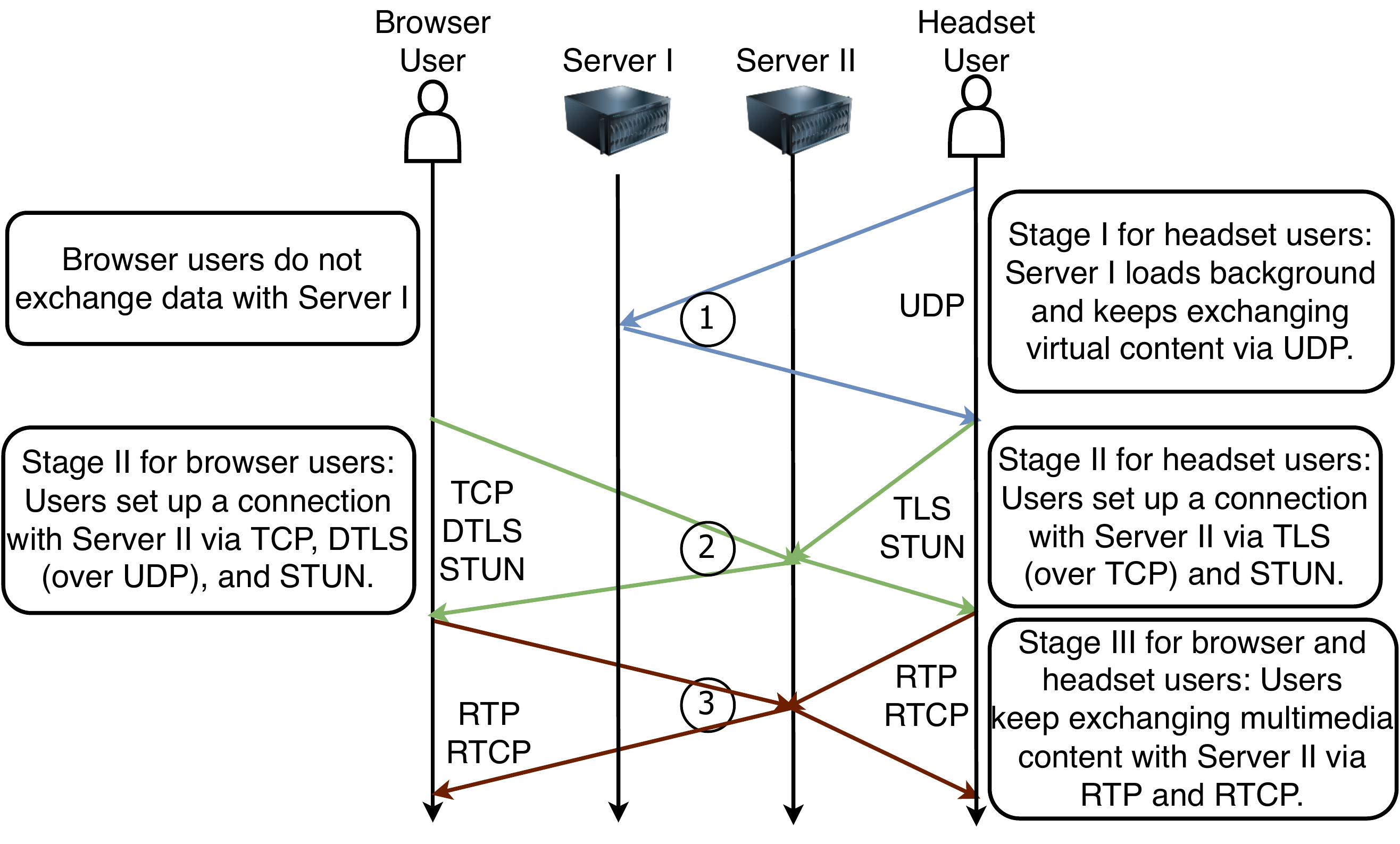}
    \includegraphics[width=0.99\columnwidth]{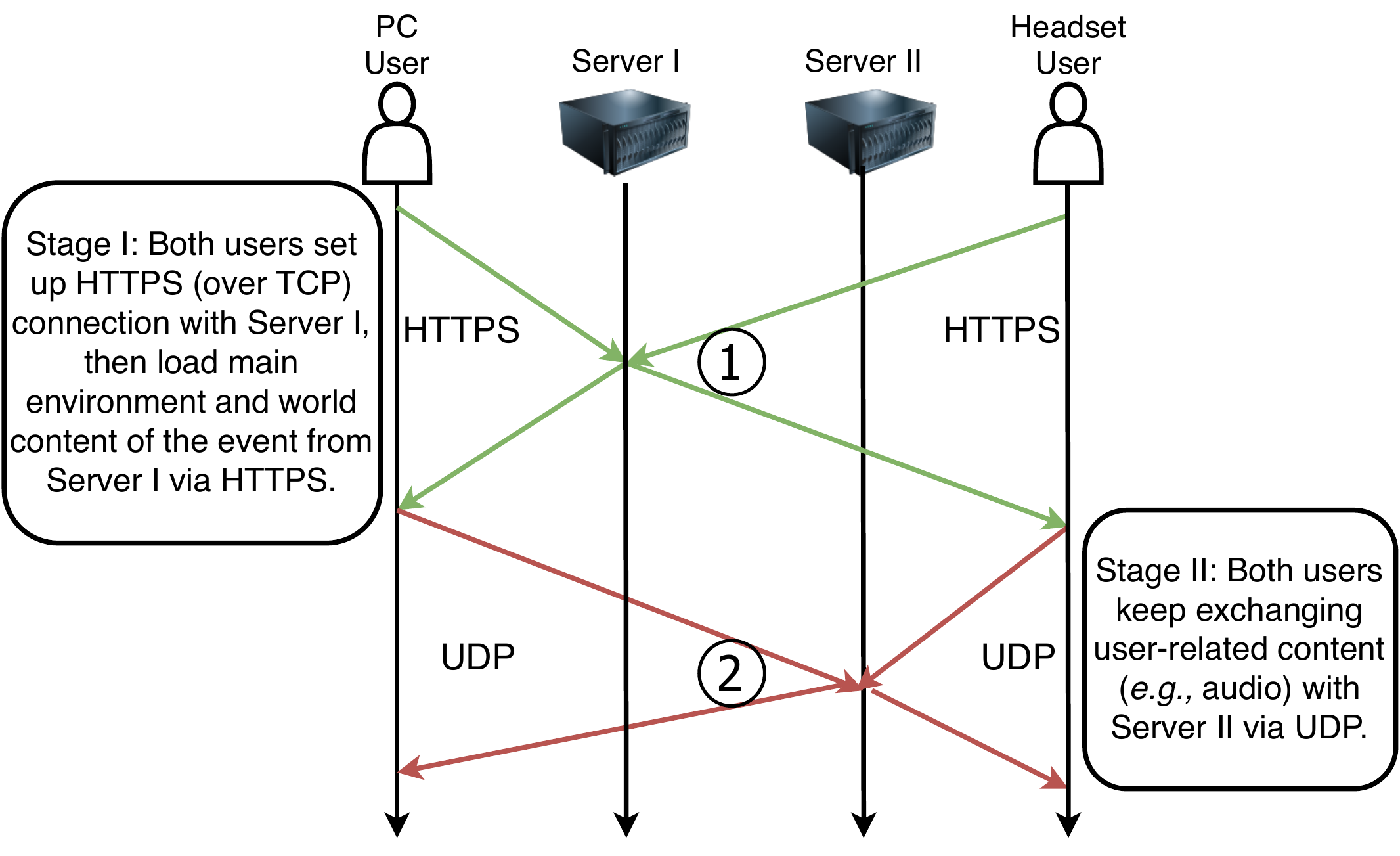}
    \vspace{-0.1in}
    \caption{\small The process of establishing connections and exchanging data between the clients and the servers for Workrooms (top) and AltspaceVR (bottom). 
    }
    \label{fig:protocol}
    \vspace{-0.2in}
\end{figure}

\begin{figure*}[t]
\centering
\subfigure[Throughput (Workrooms)]{
\begin{minipage}[t]{0.3\linewidth}
\centering
\includegraphics[width=0.99\linewidth]{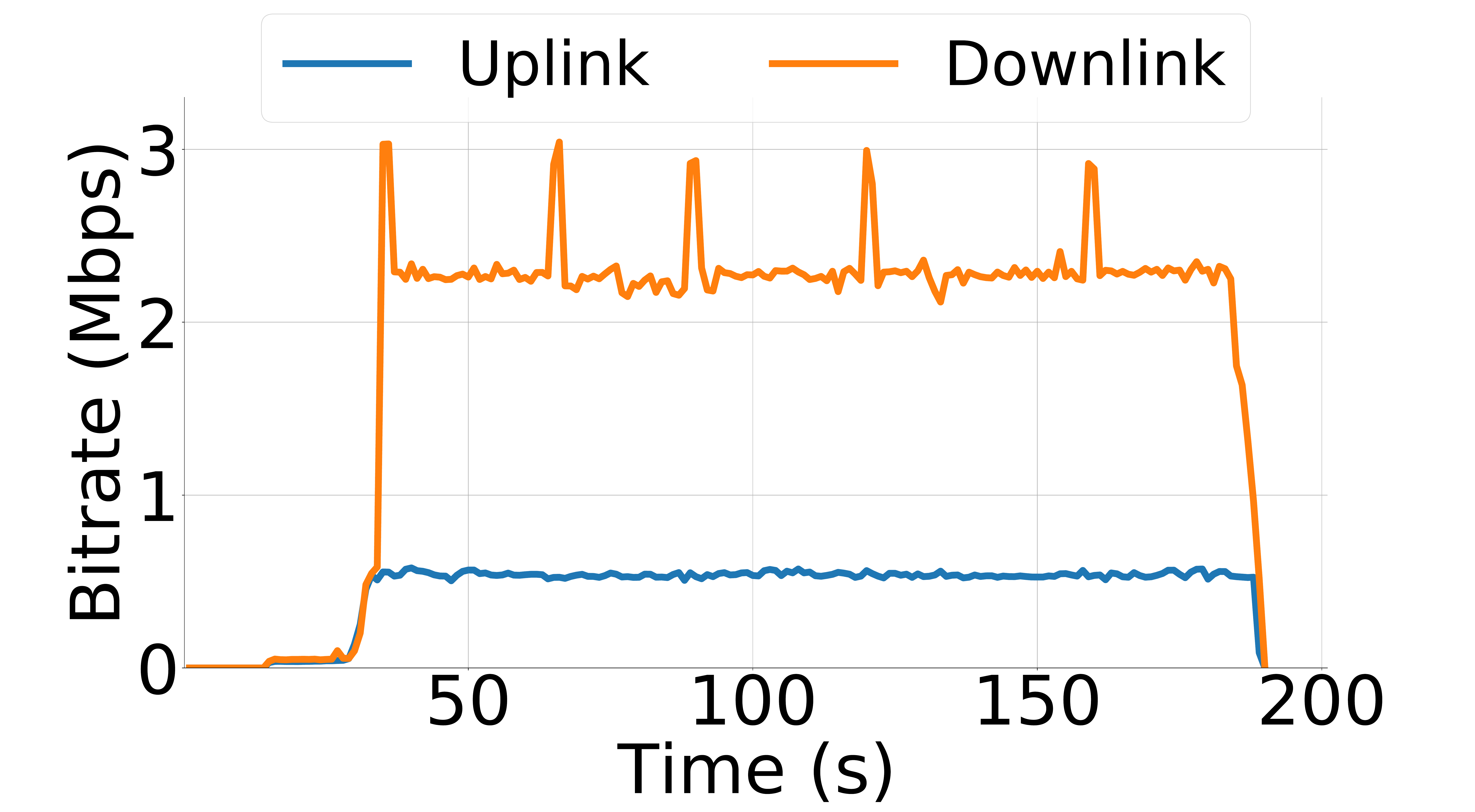}
%\caption{Throughput (Workrooms)}
\end{minipage}
}
\subfigure[Scalability (Workrooms)]{
\begin{minipage}[t]{0.3\linewidth}
\centering
\includegraphics[width=0.99\linewidth]{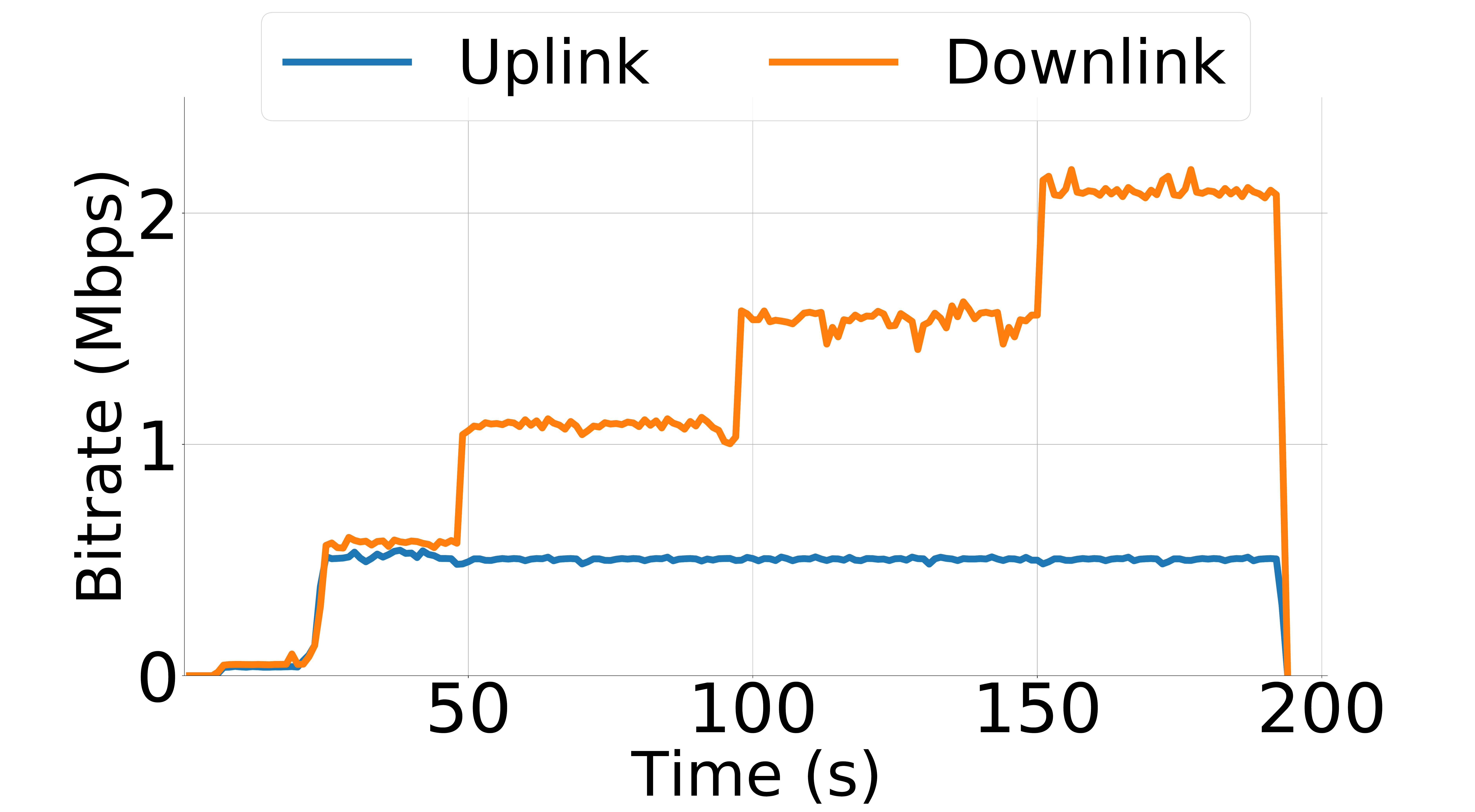}
%\caption{Scalability (Workrooms)}
\end{minipage}%
}%
\subfigure[Audio Data (Workrooms)]{
\begin{minipage}[t]{0.3\linewidth}
\centering
\includegraphics[width=0.99\linewidth]{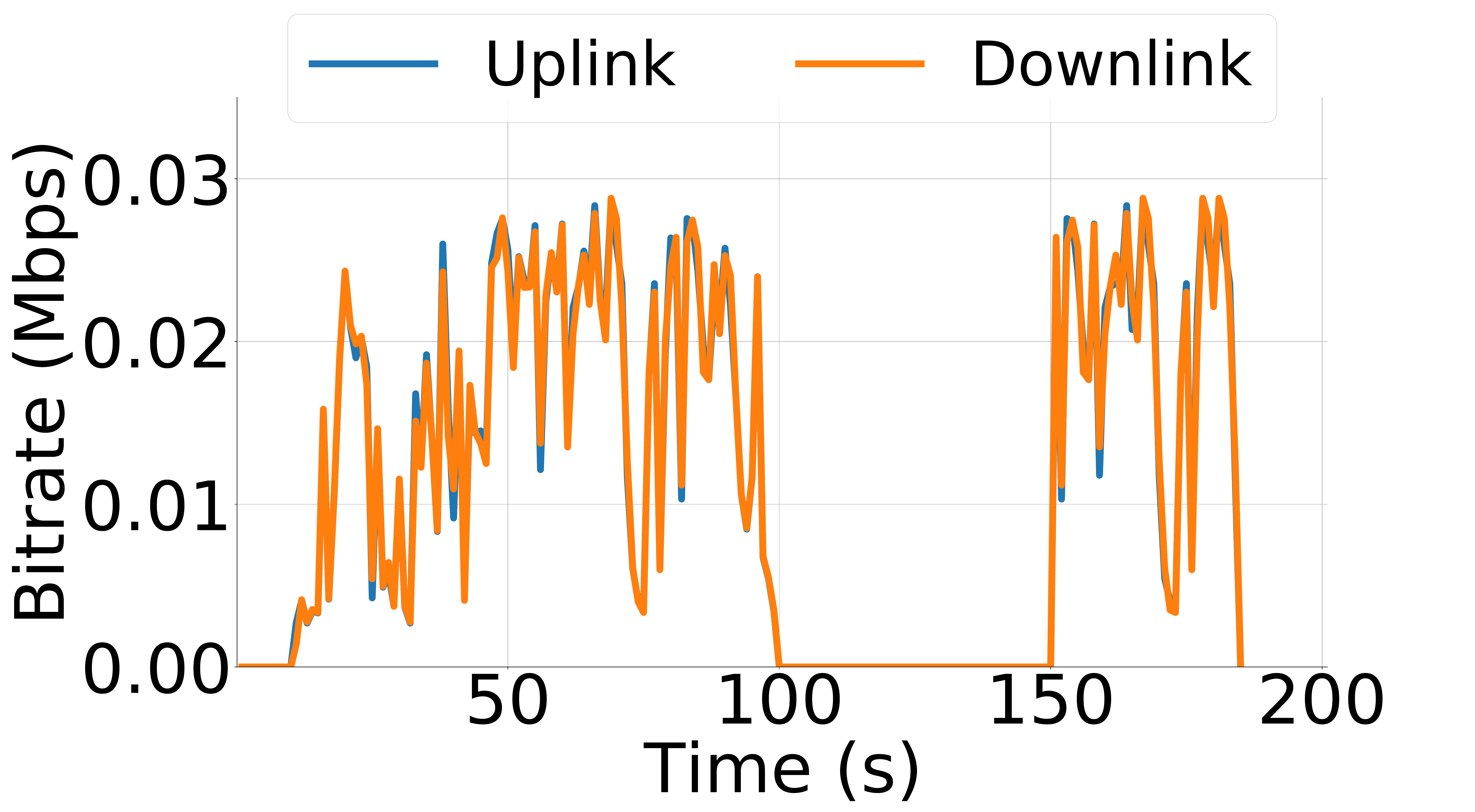}
%\caption{}
\end{minipage}
}
%\crz{the caption part shoule be in subfigure[]}
\quad

\subfigure[Throughput (AltspaceVR)]{
\begin{minipage}[t]{0.3\linewidth}
\centering
\includegraphics[width=0.99\linewidth]{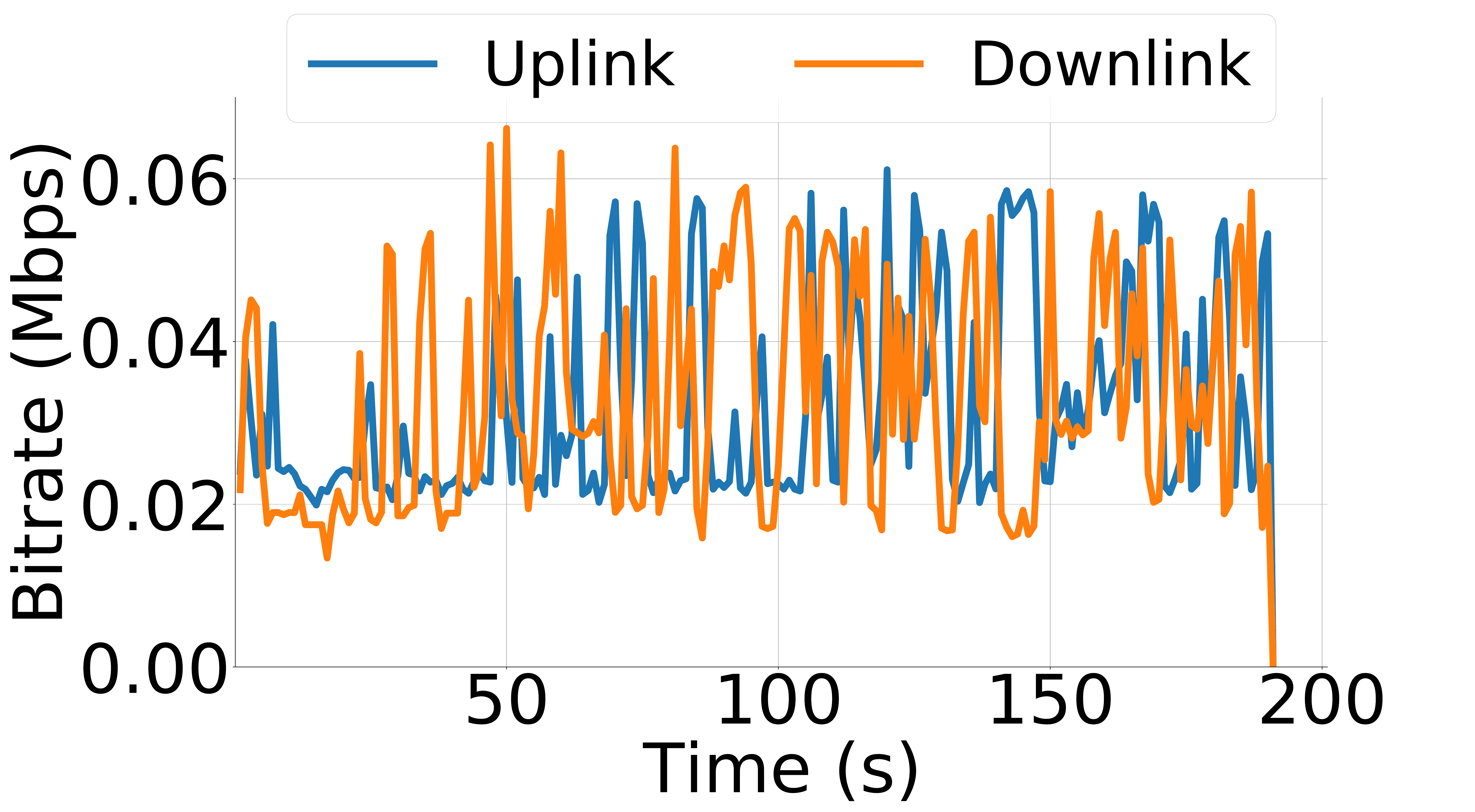}
%\caption{fig2}
\end{minipage}
}%
\subfigure[Scalability (AltspaceVR)]{
\begin{minipage}[t]{0.3\linewidth}
\centering
\includegraphics[width=0.99\linewidth]{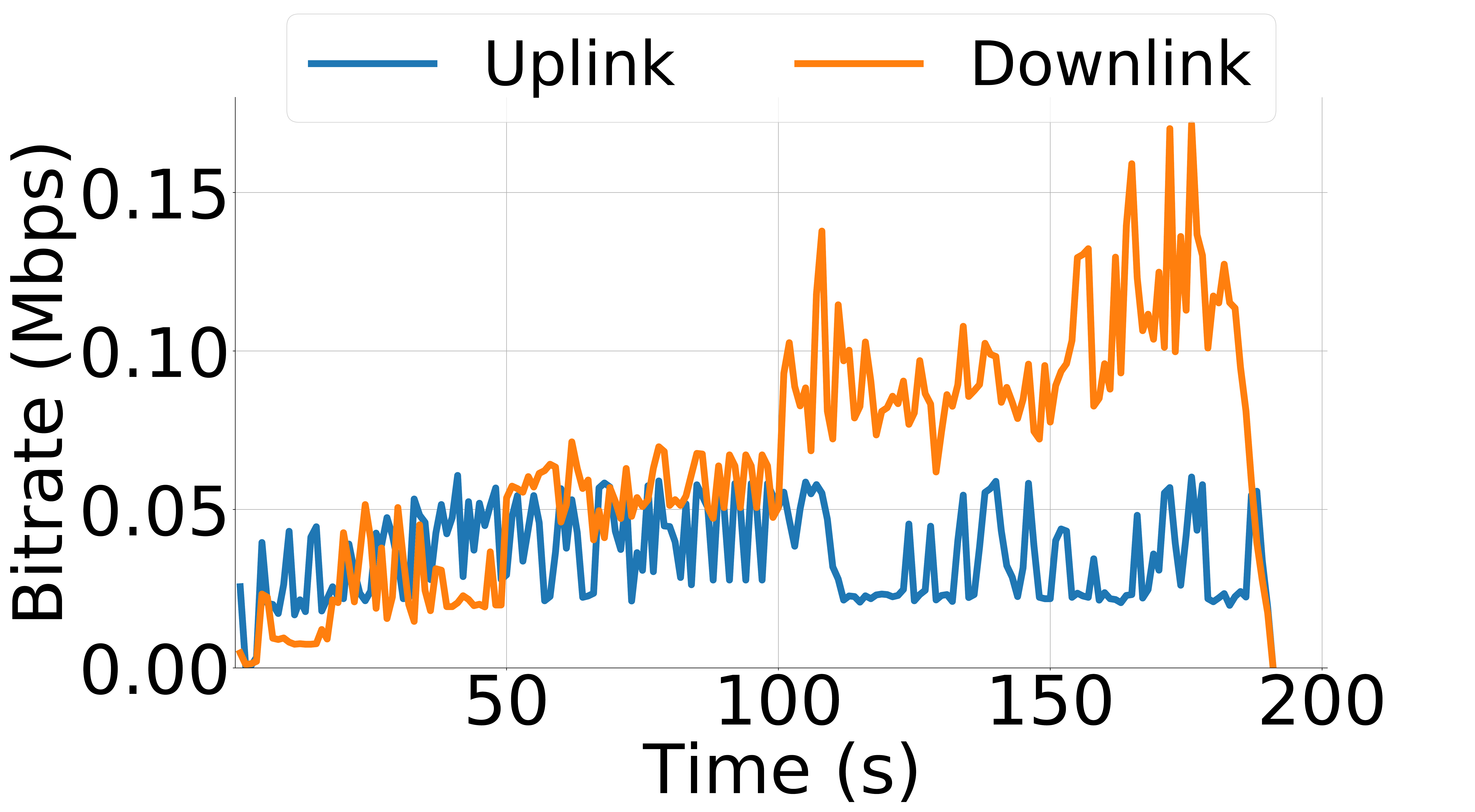}
%\caption{fig2}
\end{minipage}
}%
\subfigure[Audio Data (AltspaceVR)]{
\begin{minipage}[t]{0.3\linewidth}
\centering
\includegraphics[width=0.99\linewidth]{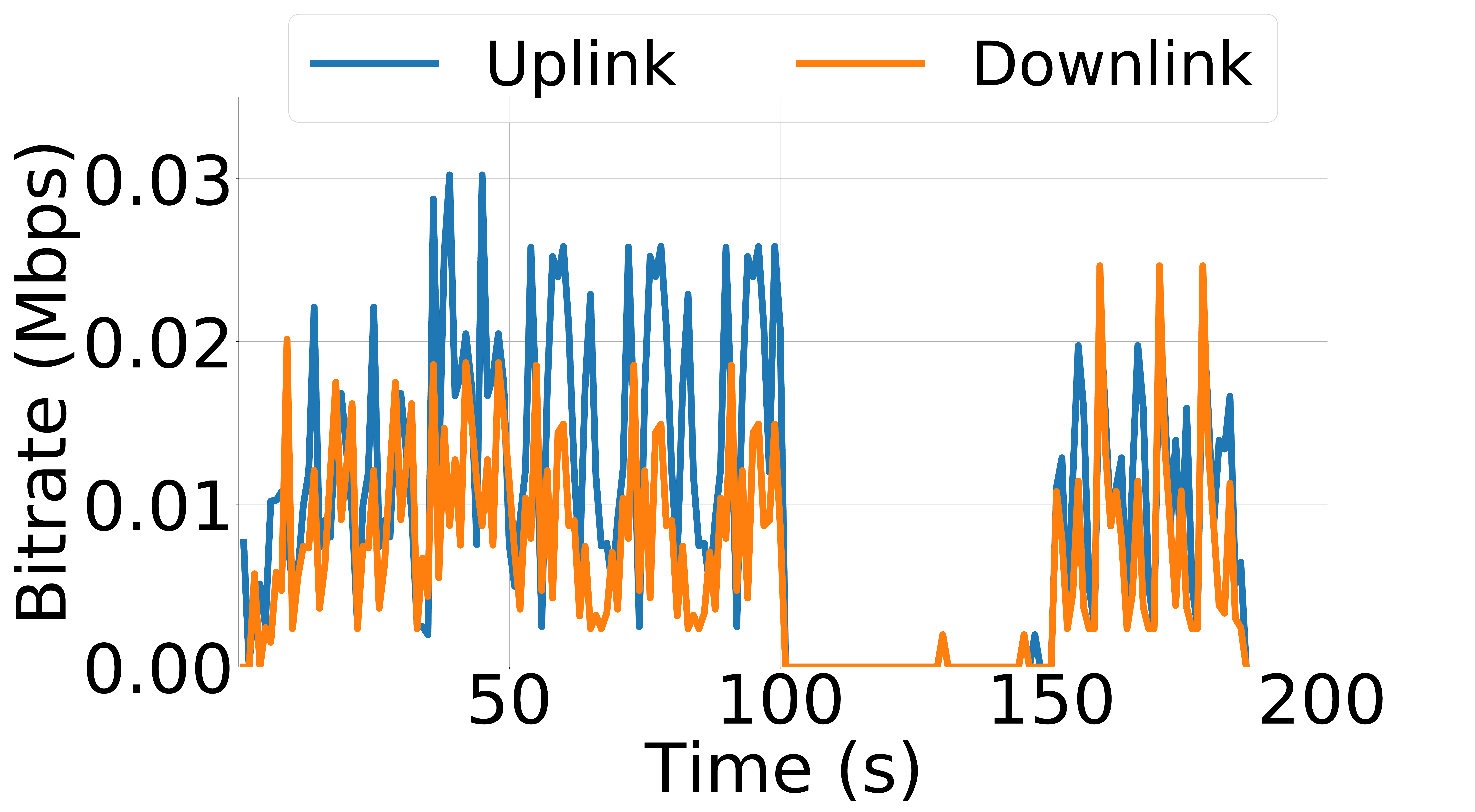}
%\caption{fig2}
\end{minipage}
}%

\centering
\caption{\rz{Comparison of throughput, scalability, audio data between Workrooms and AltspaceVR. a) and d) $U_1$'s uplink and downlink throughput of UDP flow in Workrooms %(VC flow + MM flow) 
and Altspace VR; b) and e) $U_1$'s uplink and downlink throughput of the VC flow (no change for MM flow) in Workrooms and UDP flow in AltspaceVR (three additional users $U_3$, $U_4$, and $U_5$ join at 50, 100, and 150 s, respectively); c) and f) comparison of audio data for $U_1$'s uplink and $U_2$'s downlink in Workrooms and AltspaceVR (both users mute from 100 to 150 s).} \note{(Comment 4 from Reviewer 3)}}
\label{fig:casestudy}
\vspace{-0.2in}
\end{figure*}

In this section, we conduct a reality check of the Metaverse by comparing the network operation and performance of Workrooms and AltspaceVR. %, two representative social VR platforms. 
\rz{As shown in Table~\ref{tab:social-VR comparision}, AltspaceVR is one of the earliest initiatives, and Workrooms is the most recent effort of the Metaverse,
%respectively, 
both supporting the majority of the listed features.
Thus, comparing them will give us a better understanding of the state-of-the-art of Metaverse.} \note{(Comment 3 from Reviewer 4)}
%Thus, we can get a better understanding of the current state of the Metaverse by comparing them.

%
In our previous work~\cite{cheng2022reality}, we dissected how Workrooms works. Our key findings are as follows:

\BULLET Workrooms primarily employs two servers to communicate with its clients, one is for delivering virtual content and the other is for streaming/exchanging audio and video data, as shown in Figure~\ref{fig:protocol} (top).

\rz{
\BULLET Workrooms requires $\sim$25s to initialize, by primarily performing local setup and rendering without much network activity (Figure~\ref{fig:casestudy}a and Figure~\ref{fig:casestudy}b).
} \note{(Comment 5 from Reviewer 3)}

\BULLET 
With two headset users in Workrooms, each user's downlink throughput is about 2--3 Mbps and the uplink throughput is $\sim$0.6 Mbps (Figure~\ref{fig:casestudy}a).
However, the downlink throughput linearly increases with the number of headset users, indicating that the current design of Workrooms may face scalability issues (Figure~\ref{fig:casestudy}b).

\BULLET Workrooms does not consider situations not requiring server involvement, % (\eg peer-to-peer communication), 
but simply lets the server process and forward all users' data, resulting in unnecessary communication overhead (Figure~\ref{fig:casestudy}c).

We perform the same experiments on AltspaceVR to understand the differences between the two platforms.
%
%\vspace{0.05in}
%\noindent {\bf Experiment Setup.}
We conduct a series of experiments with a 3-minute duration.
We use a Macbook Pro as the WiFi access point, which is connected to a high-speed home network via Ethernet for Internet access. 
We capture and analyze network traffic using the Wireshark packet analyzer\footnote{\url{https://www.wireshark.org/} (accessed on \accessdate)}.
%
% Since Workrooms supports user access via either Oculus Quest 2 (the only VR device for Workrooms at this moment) or Web browsers, and AltspaceVR supports user access via almost any VR headset or PC (Windows or MacOS).
% 
%We conduct most experiments with two users, U1 and U2, both using Oculus Quest 2 to access Workrooms and AltspaceVR. 
% \rz{We conduct experiments with at most five users, all of whom access Workrooms and AltspaceVR using Oculus Quest 2. }
%specified by Work-U1 and Work-U2 (or Alt-U1 and Alt-U2), respectively.
%

% \crz{For the second setting (G2), U1 uses Oculus Quest 2 to access Workrooms (or AltspaceVR), and U2 uses Google Chrome to access Workrooms (or the Windows application client to access AltspaceVR), specified by Work-G2-U1 and Work-G2-U2 (or Alt-G2-U1 and Alt-G2-U2), respectively.}
%

%To maintain uniformity in experimental metrics, our data is collected starting with the user's entry into the workroom meeting room (or event in AltspaceVR).
%
%We select the ``meeting'' event of AltspaceVR for our experiment, which has similar functionality to the workroom.

\vspace{0.05in}
\noindent {\bf Network Protocol Analysis.}
%As the first step of our study, we aim to reverse-engineer the usage of network protocols employed by Workrooms and AltspaceVR.
We first compare the network protocols employed by Workrooms and AltspaceVR.
%
%To learn how headset users connect with the server from PC users, 
Besides headsets, we use Google Chrome to access Workrooms and the Windows application to access AltspaceVR from a PC.
%
%Through multiple experiments, 
We find that users' devices communicate with two servers in both Workrooms and AltspaceVR.
%
% However, the way the user's device is connected to the two servers and the protocols transmitted are different.
%
Figure~\ref{fig:protocol} summarizes the process of establishing connections and exchanging data between the clients and the servers in Workrooms (top) and AltspaceVR (bottom).

In Workrooms, the connection with Server I starts during the loading period (\ie when the loading progress bar is displayed).
All data exchanges are over User Datagram Protocol (UDP).
We have proven that this flow transmits virtual content~\cite{cheng2022reality} and refer it to as virtual content (VC) flow.
The connection with Server II starts when users enter the meeting room.
The headset and browser clients have a slightly different way of establishing connections with Server II.
First, they both establish a Transmission Control Protocol (TCP) connection with Server II, while using Session Traversal Utilities for NAT (STUN) protocol to traverse network address translator (NAT) gateways.
%
%After that, 
The headset and Server II then transfer 1-3 Transport Layer Security (TLS, a secure communication protocol over TCP) packets to each other.
However, %after that, %establishing a TCP connection with Server II, 
the browser client does not transmit any additional TCP packets but establishes a Datagram Transport Layer Security (DTLS, a secure communication protocol over UDP) connection with Server II.
%
%DTLS is a secure communication protocol running over UDP.
%
%After the connection is established, 
Finally, both browser and headset clients use Real-time Transport Protocol (RTP) and RTP Control Protocol (RTCP) to exchange multimedia content with Server II.
%
%RTP is used to transmit multimedia streaming (\ie audio and/or video), and RTCP is used to monitor data delivery. 
\iffalse
In our previous work, we prove that the server uses Web Real-Time Communication (WebRTC)\footnote{\url{https://webrtc.org/} (accessed on \accessdate)} technology to transmit multimedia streaming for browser clients.
%
WebRTC consists of a series of application programming interfaces (APIs) and communication protocols to enable real-time communication between the server and web browsers (and/or mobile applications).
%
We speculate that the server also uses WebRTC for headset clients, although we do not have concrete evidence yet. 
\fi
We refer to this flow as multimedia (MM) flow. % (MM for short).

In AltspaceVR, however, the client-server connections work in a different way.
First, the client downloads 10-20MB of data from Server I using the Hypertext Transfer Protocol Secure (HTTPS) protocol, % \bo{(the difference comes from different events)}, %and at this time 
when the client screen displays ``downloading world content''. 
Only users who join the event for the first time need to download the data.
Then, the client downloads 300-500KB of data from Server I via another HTTPS connection, %and at this time 
when the client screen displays ``loading main environment''. 
%
%After loading the main environment, the client and server establish a new HTTPS connection every 2-3 seconds, with approximately 6000 bytes and 2000 bytes for the downlink and uplink, respectively.

The connection with Server II starts when users finish loading.
All data exchanges are over UDP.
%
\iffalse
The first 3 bytes of the UDP payload may be used to distinguish the data source.
%
Packets sent by the server are all assigned 0, and those sent by the client are assigned randomly.
%
The fourth byte is used to distinguish the data type, such as audio data. 
%We find that 0x01 may be related to user authentication information, and 0x04 and 0x05 are related to the audio data of users (Section xxx).
%
Bytes 5 to 8 are used to record the time difference between the current received packets and the first received packets of the connection.
%
Bytes 9 to 12 distinguish between different data streams, and the contents of this part are the same for all packets transmitted in a connection between the server and a client.
\fi
%
Since this UDP flow is the only flow %that is continuously transmitted 
after users enter the event, we speculate %that this flow 
it is for %transmitting 
user-related content.
Through further analysis, we find that this UDP flow follows a custom protocol.
The fourth byte of the UDP payload is used to distinguish the data type, such as audio data.

\vspace{0.05in}
\noindent {\bf Network Performance.}
Next, we compare the network performance of two platforms based on key findings of Workrooms.
\rz{The scalability experiments involve up to five users (denoted as $U_{i}$), all with Oculus Quest 2, and all other experiments have only two users (\ie $U_1$ and $U_2$).} \note{(Comment 4 from Reviewer 3)}
Figure~\ref{fig:casestudy}d shows the throughput (\ie bitrate) of the UDP flow in AltspaceVR with two users.
We find that since users have downloaded the event content %from the server 
before entering the event, the throughput (less than 0.06 Mbps) after that %entering the event 
is much smaller than Workrooms.
%
%We also observe that when entering the event, the event creator uploads 13KB data (not shown) to ports 5055 and 5056 of the server, respectively.
%
%In contrast, other users do not have this data stream.
%
%We suspect that this 13KB data is related to the authentication information of the event creator. 
%(not confirmed yet).
%
Figure~\ref{fig:casestudy}e shows the throughput of the UDP flows in AltspaceVR, where we let three other headset users join the experiment at 50, 100, and 150 s, respectively.
We find that AltspaceVR also faces scalability issues, with the downlink bitrate %\bo{roughly doubling} 
increasing almost linearly every time a new user joins ($\sim$0.03Mbps). 
However, its increase is much smaller than Workrooms ($\sim$0.5Mbps). 

Figures~\ref{fig:casestudy}c and~\ref{fig:casestudy}f show the comparison of the audio data for $U_1$'s uplink and $U_2$'s downlink with two users. % in Workrooms and AltspaceVR.
%
%As we mentioned before, the fourth byte of the UDP payload in AltspaceVR is used to distinguish the data type.
%
%We mute the user and find that the UDP flow with the fourth byte is 0x04 or 0x05 is affected, so we speculate that these two types of this UDP flow are related to the user's audio data.
%
In Workrooms, we observe that the bitrate on the downlink of $U_1$ exactly matches that of the uplink of $U_2$, and vice versa (Figure~\ref{fig:casestudy}c).
This indicates that the server simply forwards one user’s audio data to others without further processing. 
%
%However, given that the users are all in the same subnet, it is feasible to directly perform peer-to-peer (P2P) communication without the server forwarding. 
%
%Workrooms currently does not employ P2P techniques to reduce the bandwidth overhead.
%
However, in AltspaceVR, the uplink audio data of one user does not exactly match the downlink audio data of another user, indicating that the server processes the audio data before forwarding it (Figure~\ref{fig:casestudy}f).
Also, most of the time, the downlink throughput of a user is lower than the uplink throughput of the other user, which indicates that the server may optimize the audio data uploaded by users.%, such as compression.
%
%We also find that during user muting, both uplink and downlink occasionally send a packet with 1984 bytes data. 
%
%We do not know the cause of this phenomenon at this moment.

\vspace{0.05in}
\rz{
\noindent {\bf Accessibility and Security \& Privacy Issues.} Finally, we study the accessibility and potential security and privacy issues, considering that these are the key requirements to the success of the Metaverse (\S\ref{sec:definition}).

\BULLET  AltspaceVR supports nearly all VR headsets, whereas Workrooms supports only the Oculus Quest 2 headset.

\BULLET As shown in Table~\ref{tab:social-VR comparision}, neither AltspaceVR nor Workrooms offers shopping/NFTs. Hence, there are currently limited concerns regarding the security of transactions.
However, with the active participation of high-tech companies such as Meta\footnote{\url{https://bit.ly/3zc59gc} (accessed on \accessdate)} in the NFT, these concerns may emerge in the near future.
In addition, AltspaceVR provides the personal space that protects users from harassment, but Workrooms does not.
}
\note{(Comment 1 from Reviewer 3)}

To summarize, by comparing Workrooms and AltspaceVR, we have the following findings.

\BULLET %Workrooms and AltspaceVR follow different transmission patterns.
%
%Workrooms does not require the user to download data in advance, but rather transmits data continuously during the users' access, resulting in higher bandwidth consumption.
%
AltspaceVR requires users to download event data in advance, and transfers only user-related data after that, requiring less bandwidth consumption than Workrooms.

\BULLET Both platforms %Workrooms and AltspaceVR 
face scalability issues, 
%
%However, 
although AltspaceVR does not cause significant bandwidth consumption ($\sim$0.18 Mbps for downlink %bandwidth consumption 
with five users). % join the event).

\BULLET Unlike Workrooms, AltspaceVR processes the data uploaded by users before forwarding it,  reducing the size of data received by other users. %\crz{spatial sound?}.

%% file: 06.discussion.tex
In this section, we %introduce other types of Metaverse platforms, 
discuss the technical and design challenges of building the Metaverse and point out opportunities for further innovation.
%
%Implementing a scalable, secure, and reliable network system with high QoE is crucial to the success of the Metaverse.
%
%It is foreseeable that the Metaverse will pose a huge challenge to the underlying networks. % technology. 
%\bo{what is metaspace here?}
%
%Even existing 5G technology may not be able to solve it.

\noindent {\bf Technical Challenges.} First, the operation of the Metaverse will generate a large amount of data, such as metadata created by sensors, a shared virtual space for social activities of users, and the transmission of high-resolution video streams, requiring a huge network bandwidth.
However, the existing 5G technology may not be sufficient to support the Metaverse.
%
%\rz{On the one hand, given that the future Metaverse may host thousands of millions of users concurrently in the shared virtual environments, the scalability issue will cause it to explode in bandwidth demand.}
%
\rz{%On the other hand, although 
As shown in Figure~\ref{fig:casestudy}, the throughput required for current social VR platforms is low.
This is because the avatar of the majority of existing social VR platforms has only the upper torso, and the movement of the  avatar is not driven by the actual movement of the user but operated by the hand-held controllers of the VR headset.
%~\cite{tanenbaum2020make}. 
%
However, the future Metaverse will necessitate high-quality full-body avatars to provide a truly immersive experience for millions of concurrent users in the shared virtual environment, demanding higher throughput than we have observed on today's social VR platforms. 
For example, existing work such as Holoportation~\cite{orts2016holoportation} reveals that the bandwidth required to deliver a photo-realistic 3D model of a human body by capturing its motion in real-time %for a single user 
can exceed 1 Gbps.} 
\note{(Comment 3 from Reviewer 3)}
%\csq{add a citation here}\rz{This is also the  ~\cite{orts2016holoportation}}
%

%
% Although 5G can reach a maximum throughput of, in theory, 10-20 Gbps, considering the scalability demand, the bandwidth requirement of Metaverse may exceed what 5G can offer. %~\cite{lee2021Metaverse}.
%

\rz{To this end, we discuss several potential solutions to address the scalability issue. Viewport-adaptive optimizations, which aim to deliver only the content visible to users for saving bandwidth, can alleviate the scalability issue. 
However, if enormous avatars are visible in the user's viewport, the required network bandwidth to transmit their data and resource utilization on the device for rendering may still be high.
Moreover, the server needs to predict the future viewport of the user to determine the to-be-delivered content, which may negatively impact the user experience if the prediction is inaccurate~\cite{zhang2021deepvista}}.
\rz{Another potential direction is peer-to-peer (P2P) communication techniques.
In P2P, user devices will need to combine the content received from multiple parties and then render the virtual world accordingly. %that everyone is accessing. 
However, given that rendering is still performed on the client-side, the on-device resource consumption could be excessive.
%, resulting in poor rendering quality.
}

\rz{Another promising strategy to address the aforementioned scalability issue is to utilize remote rendering~\cite{koller2004protected}, in which the server is responsible for performing rendering tasks. 
In this scenario, even though there are a significant number of concurrent users (especially when their avatars are clustered together), the servers will render the entire scene in a user's viewport %, including visible avatars, 
into a 2D video frame. Hence, the amount of transmitted data is independent of the number of users, alleviating the scalability issues.
Nevertheless, remote rendering still poses technical challenges. For instance, similar to
viewport-adaptive optimizations, the performance of remote rendering depends on the accuracy of viewport prediction.
Additionally, the server may have to render the same number of scenes as the number of users since different users may have different viewports.}
\note{(Comment 2a from Reviewer 1)}
%, and it takes time to transmit to the VR headset over the network, which may cause high network latency, as discussed in the following. }

Second, network latency is critical to the user experience.
Given that users may access the Metaverse from different parts of the world, ensuring low latency when users are across geographically distributed regions is a practical challenge.
Meanwhile, sensors in the Metaverse, such as those on XR headsets and haptic devices, require latency as low as tens of milliseconds to maintain an immersive user experience~\cite{zhang2022sear}.
Similar to the motion-to-photon latency in VR, in the Metaverse the latency between the motion of a user and its reflection perceived by others is a key metric to optimize.
%
%\note{ Comment 2b from Reviewer 1.}\todo{Remote rendering~\cite{koller2004protected} of virtual content at the network edge is a promising direction for reducing the above latency.}
%
%Although %\bo{high-band} the latency of mmWare 5G can theoretically go down to one  millisecond, as we discussed in Section~\ref{sec:definition}, it is challenging to deploy and apply at a large scale.

Third, the security and privacy issues in the Metaverse deserve our attention.
%
%In our measurement study, we find that currently commercial social VR platforms widely use secure communication protocols, such as TLS or DTLS, to protect transmitted data.
%
Although commercial social VR platforms employ secure communication protocols (\eg TLS and DTLS) to protect transmitted data, as verified in our measurement study, the Metaverse may still lead to many security concerns, such as users' identification information.
Since it requires users to access with headsets, they %users in the Metaverse 
often need to identify themselves %their personal information 
with biometric information, which could be a target of security attacks~\cite{mathis2021fast}.
%
%However, the user's biometric information can be attacked by \crz{based on the citation of requirement #3}~\cite{xxx}.
%
Digital twins in the Metaverse also need proper protection.
There will be a large number of complex ML models for supporting digital twins, which in turn influence %and guide 
objects in the physical world.
If these models %in the digital twin 
are attacked, there will be unpredictable consequences in the physical world.
Moreover, user-worn headsets will continuously collect personal information (\eg biometric information and user behavior) and transmit it to the server for training the model, which causes privacy concerns.
%
%to improve the QoE\rz{, such as training ML models}, which results in privacy concerns.
%
Storing biometric data and digital twins of the Metaverse in the blockchain is a possible direction~\cite{ryskeldiev2018distributed}.

\iffalse
\rz{Another approach is to leverage federated learning (FL) to train the ML models in the Metaverse.
%
FL is a distributed ML method that utilizes users' local data to train a global model by iteratively communicating model parameters between participating users and the server~\cite{li2020federated}.
%
It can train ML models using localized data from isolated data sources, such as mobile devices.
%
Thus it becomes a feasible method for training ML models in the Metaverse without compromising user privacy.
%
Utilizing FL in the Metaverse, however, presents several technical challenges.
%
For example, the computing power of VR headsets may not be sufficient for FL training, which could negatively impact the user
experience. Battery drainage is another concern.
%such as rapidly draining the battery.
%
Moreover, clients may present heterogeneous resource and data issues~\cite{li2020federated}, which could affect model performance and increase communication overhead.
%
Another challenge is that in the FL scenario, each client holds only positive label data, which may lead to scalability issues of the model~\cite{yu2020federated}.}
\fi

\rz{Besides data privacy, harassment is another emerging concern in the Metaverse.
The Metaverse has not only text-based or voice-based harassment, which has been studied for traditional social media, but also body movement-based harassment that can reflect the users' movement via their avatar through various sensors. 
However, the protection mechanism against this type of harassment has not been thoroughly investigated.
As shown in Table~\ref{tab:social-VR comparision}, several platforms have implemented the personal space feature.
Nevertheless, this feature is a passive defense against harassment.
Since it restricts social interaction, users may elect to enable it only after harassment happens.
%
%However, by then, the harassment has already occurred.
%
Therefore, a more desirable mechanism should be able to detect potential harassment prior to its occurrence without impacting the user experience.}
\note{(Comment 2b from Reviewer 1)}

\noindent {\bf Real-world Challenges.} \rz{In addition to the above technical challenges, we also need to consider the following issues related to the physical world when designing the Metaverse.}
\rz{First, the Metaverse %is a liberal space, but this 
may cause ethical concerns. 
For example, it allows users to freely choose their avatars, but not all avatars are equally in demand. 
According to a study, users have a low demand for dark-skinned and female avatars\footnote{\url{https://www.thenifty.com/race-and-nfts-636/} (accessed on \accessdate)}, raising issues about race and gender representation in the Metaverse.} %\csq{what is the potential solution?}
\rz{
Second, as the Metaverse becomes commonplace in our daily lives, user addiction will be a crucial issue\footnote{\url{https://bit.ly/3RWEvzc} access on \accessdate}. \note{(Comment 5 from Reviewer 4: Add a footnote.)}
People may %Users are likely to use 
rely on the Metaverse to escape from the real world, as described in the novel {\em Snow Crash}.
%
%The COVID-19 pandemic seems to facilitate this process\footnote{\url{https://bit.ly/3PwA781} (accessed on \accessdate)}.
%\note{Comment 5 from Reviewer 4. Change the footnote}
%
%Recent surveys show that 51\% of U.S. adults use social media at a higher rate during the COVID-19 pandemic\footnote{\url{https://bit.ly/3q8neXW} (accessed on \accessdate)}.
%
Beyond better regulation and guidance, how to effectively address this issue is still an open problem.
Finally, virtual crimes in the Metaverse deserve our attention.
The transactions in the Metaverse are conducted through blockchain-based NFTs and cryptocurrencies.
Decentralization and non-regulation are the two main features of blockchain, which are prone to crime.
In 2021, the worth of criminal activity regarding cryptocurrencies was up to \$14 billion\footnote{\url{https://reut.rs/3s0yEwC} access on \accessdate}.
Since the Metaverse is a decentralized and free virtual world, our efforts to guide and monitor these issues in the real world may not be replicable in it.
%
%Therefore, 
How to effectively address the above issues in the Metaverse still deserves in-depth study.}
%\note{delete the sentence ``Beyond better ... problem''?}
\note{(Comment 1 from Reviewer 2)} %and Comment 5 from Reviewer 4)} 

%
%In addition to better regulation and guidance, a practical solution is to make modules in the Metaverse, such as digital twins, work better to positively impact the real world, making addicted users aware that the real world has more value than the Metaverse.

\iffalse
Finally, Virtual crimes in the Metaverse also deserve our focus.
%
The transactions in the Metaverse are conducted through blockchain-based NFTs and cryptocurrencies.
%
Decentralization and non-regulation are the two main features of the blockchain, but they are also prone to crime.
%
In 2021, the worth of criminal activity regarding cryptocurrencies is up to \$14 billion\footnote{\url{https://reut.rs/3s0yEwC} access on \accessdate}.
\fi

%\bo{security, privacy, user addiction, and virtual crime.} 
%``our vision'' part? \bo{that is about the requirements. here we should summarize the challenges based on the measurement results.}
%}

% \bo{expand this paragraph a bit?}
% \crz{paragraphs: letency, security, acces ..}
% %\crz{latency:IMC, sync}

%\vspace{0.05in}
%\noindent {\bf Opportunities and Future Directions.}

%Information privacy, user addiction, and virtual crime.

%% file: 07.conclusion.tex
While the Metaverse has been deemed as the NextG Internet, much of the discussion, in both industry and academia, has focused on %the vision and 
its potential. 
%
%Admittedly, Metaverse could be built on top of the fast-developing 5G, AR/VR/MR, blockchain, cryptocurrencies, HCI, \etc
%
\rz{In this article, after reviewing the current hype in the industry, %such as 5G, AR/VR/MR, edge computing, blockchain, cryptocurrencies, machine learning, and HCI, 
we present the definitions of Metaverse, its enabling technologies, and our vision of its technical requirements.
%
%After that, 
We then introduce existing social VR platforms that can be viewed as early prototypes of Metaverse.} \note{(Comment 1 from Reviewer 4)}
By measuring and comparing two representative social VR platforms, Workrooms and AltspaceVR, we point out the technical challenges and opportunities for future development. % of the Metaverse.
Given its multidisciplinary nature~\cite{park2022metaverse}, we hope to see more initiatives emerging from not only the networking research community, but also other related disciplines such as social sciences, economics, computer graphics, AR/VR/MR, HCI, security, and privacy.